# Electron-Hole Asymmetry in Single-Walled Carbon Nanotubes Probed by Direct Observation of Transverse Quasi-Dark Excitons

Yuhei Miyauchi[1,2], Hiroshi Ajiki[3] and Shigeo Maruyama[4,*]

[1]*Institute for Chemical Research, Kyoto University, Uji, Kyoto 611-0011, Japan*
[2]*Departments of Physics and Electrical Engineering, Columbia University, New York, New York 10027, USA*
[3]*Photon Pioneers Center, Osaka University, 2-1 Yamadaoka, Suita, Osaka 565-0871, Japan*
[4]*Department of Mechanical Engineering, 7-3-1 Hongo, Bunkyo-ku, Tokyo 113-8656, Japan*

**Abstract**

We studied the asymmetry between valence and conduction bands in single-walled carbon nanotubes (SWNTs) through the direct observation of spin-singlet transverse dark excitons using polarized photoluminescence excitation spectroscopy. The intrinsic electron-hole (e-h) asymmetry lifts the degeneracy of the transverse exciton wavefunctions at two equivalent K and K' valleys in momentum space, which gives finite oscillator strength to transverse dark exciton states. Chirality-dependent spectral weight transfer to transverse dark states was clearly observed, indicating that the degree of the e-h asymmetry depends on the specific nanotube structure. Based on comparison between theoretical and experimental results, we evaluated the band asymmetry parameters in graphene and various carbon nanotube structures.
PACS: 71.20.-b, 71.35.Cc, 78.67.Ch

Unlike in conventional solids, electrons and holes described by the Dirac Hamiltonian in graphene [1] have linear energy-momentum dispersions with an electron-hole (e-h) symmetry known as the Dirac cone. Single-walled carbon nanotubes (SWNTs)—essentially rolled-up graphene—inherit the graphene dispersion relations with reservation of quantum confinement in the circumferential direction, identified with the wrapping vector ($n$, $m$) [2]. Hence, electrons and holes in the 1D subbands on the Dirac cone in SWNTs also exhibit e-h symmetry, which has been experimentally confirmed in the regime of a few electrons or holes [3]. Various exotic physical properties originating from the above anomalous electronic structures make SWNTs and graphene promising materials for exploring fundamental physics in low dimensional systems and for many potential applications [2, 4].

The above description of the electronic structure of graphene and SWNTs is, however, no longer true at higher energies. Away from the Fermi level, theory predicts that the band structure gradually becomes asymmetric due to overlap of electron wavefunctions and/or hopping between adjacent carbon atoms [2, 4-8]. This band asymmetry is not only quite fundamental, but is also practically important for considering future applications of carrier-doped SWNTs [7, 9] such as ultra-efficient photovoltaic devices using extremely efficient multiple e-h generation in SWNT-diodes [10]. Despite this importance, no experimental study has clarified the magnitude of the intrinsic asymmetry in graphene and its dependence on the specific chiral structure of SWNTs. There are a few transport studies that have reported strong e-h asymmetry in graphene [11-14], however metal contacts used in transport measurements [12] and/or the existence of charged impurities [13-15] might easily induce an extrinsic e-h asymmetry to the transport properties. Recent infrared optical measurements of graphene samples with high mobility have revealed near-symmetric velocities [16], suggesting the intrinsic e-h asymmetry is still small at infrared energies. Therefore, non-invasive observation such as optical spectroscopy at higher energies is essential to probe the intrinsic e-h asymmetry in graphene and SWNTs.

The optical transitions in semiconducting SWNTs are dominated by strongly bound electron-hole states called excitons [17-21]. As shown in Fig. 1(a), longitudinal excitons consist of electrons and holes in the same 1D subband ($E_{11}$, $E_{22}$, …). On the other hand, transverse excitons [21, 22] have quasi-angular momentum connecting the electron and hole states across these subbands ($E_{12}$ and $E_{21}$). Therefore, transverse excitons include information about the e-h asymmetry. This enables us to probe the intrinsic e-h asymmetry by noninvasive optical measurements. As shown in Fig. 1(b), the degenerate exciton states near the K and K' points in momentum space are theoretically predicted to yield optically active (bright) and inactive (dark) exciton states through the intervalley Coulomb interaction for both longitudinal and transverse excitons [21]. Only longitudinal dark states [23-25] have been experimentally confirmed using the Aharonov-Bohm effect [26] in strong magnetic fields.

In this Letter, we demonstrate experimental evidence of the intrinsic e-h asymmetry in SWNTs through the first direct observation of transverse dark excitons. Our observation shows novel weak transverse exciton absorption peaks approximately 200-300 meV below the lowest optically active transverse exciton peaks in polarized PL excitation (PLE) spectra. These peaks are attributed to transverse dark states that acquire finite oscillator strength (*quasi*-dark states) due to the degeneracy lifting of $E_{12}$ and $E_{21}$ transverse excitons in the K and K' valleys caused by the intrinsic e-h asymmetry in SWNTs. We found a clear (*n*, *m*) dependence in the relative intensities between bright and *quasi*-dark states. In addition, we have confirmed that these experimental results are consistent with theoretical calculations of transverse excitons taking the depolarization effect into account. From the comparison between experimental and theoretical results, we evaluated the band asymmetry parameters for graphene and SWNTs with specific chiral structures.

For the optical measurements, SWNTs synthesized by the HiPco method were dispersed in $D_2O$ with 0.5 wt % sodium dodecylbenzene sulfonate (SDBS) by vigorous sonication with an ultrasonicator for 1 h at a power flux level of 460 W/cm$^2$ [27]. These suspensions were then centrifuged for 1 h at 386 000 g and the supernatant, rich in isolated SWNTs, was used for PL measurements. The use of such ensemble samples enabled us to probe weak transverse quasi-dark exciton absorption. Near-infrared (NIR) PL emission from the sample was recorded while the excitation wavelength was scanned from 730 to 1000 nm. A CW Ti:sapphire laser (100 mW/cm$^2$) was used for the excitation, and the excitation power was monitored and kept constant during the measurements. The emission spectral slit width was 5 nm and scan steps were 5 nm on the excitation axis. NIR polarizers were placed behind the excitation laser and before the emission monochromator, respectively. The alignment of the polarizers was examined by observing the polarization of scattered light from dilute colloidal silica in water. Polarized PLE spectra were obtained with the emission polarizer oriented parallel to ($I_{VV}$) or perpendicular to ($I_{VH}$) the direction of the vertically polarized excitation, and the pure component for parallel ($I_{//}$) and perpendicular ($I_\perp$) dipoles relative to the nanotube axis were obtained using the relationships [28] $I_{//} = [(r_{\exp} - r_\perp)/(r_{//} - r_\perp)](I_{VV} + 2I_{VH})$ and $I_\perp = [(r_{//} - r_{\exp})/(r_{//} - r_\perp)](I_{VV} + 2I_{VH})$, where $r_{\exp} \equiv (I_{VV} - I_{VH})/(I_{VV} + 2I_{VH})$ is the anisotropy, and $r_{//}$ and $r_\perp$ are the maximum and minimum anisotropies for parallel and perpendicular dipoles, which are related by $r_\perp = -0.5 r_{//}$ [29]. A value of $r_{//} = 0.36$ was determined from the maximum value of observed anisotropy for SWNTs in the lower excitation energy range without any peak structure. Details of the experimental technique are presented in Ref. [28]. The validity of the spectra obtained by this technique has been confirmed by direct observation of $I_{//}$ and $I_\perp$ spectra by single nanotube spectroscopy [30].

Figure 1(c, d) shows PLE maps for parallel and perpendicular excitations to the nanotube axis. For perpendicular excitation, the observed PLE peak positions were completely different from those for parallel excitation. Both peak positions and PLE spectral shapes of the dominant peaks for parallel and perpendicular incident light are consistent with those in previous measurements [28, 30, 31]. For parallel excitation, near-infrared PL due to the e-h recombinaiton of longitudinal excitons within the first subband $E_{11}$ followed by excitation within the second subband $E_{22}$ are observed. On the other hand, $E_{11}$ PL followed by excitation of transverse excitons between the first and second subbands ($E_{12}$ and $E_{21}$) is observed for perpendicular excitation. The transverse exciton peaks have broad absorption peaks and intensity tails to the high-energy side, as has been reported previously [28, 30]. Superposition of the $E_{12}$ and $E_{21}$ exciton wavefunctions in the K and K' valleys produces the bright and dark transverse exciton states as introduced above. Hereafter, we refer the bright transverse exciton states as $E_{12}^{(+)}$.

On the low-energy side of $E_{12}^{(+)}$ we found weak but distinct absorption peaks approximately 200-300 meV below the $E_{12}^{(+)}$ absorption peaks, as shown in the outlined region in Fig. 1(d). We attribute these small absorption peaks below $E_{12}^{(+)}$ to exciton absorption by even-parity dark excitons that acquire finite oscillator strength due to the degeneracy lifting of $E_{12}$ and $E_{21}$ excitons originating from the intrinsic e-h asymmetry in SWNTs. We hereafter refer these exciton transitions as $E_{12}^{(-)}$.

Figure 2(a) shows the observed excitation energies for longitudinal and transverse excitons. $E_{12}^{(+)}$ (squares) and $E_{12}^{(-)}$ (triangles) were observed between $E_{11}$ and $E_{22}$ (circles). $E_{12}^{(+)}$ are close to $E_{22}$, while $E_{12}^{(-)}$ are relatively close to $E_{11}$. Note that phonon-related features such as phonon sidebands and Raman scatterings can be ruled out as an explanation for the $E_{12}^{(-)}$ peaks because the $E_{12}^{(-)}$ peaks do not have constant energy difference from $E_{11}$, $E_{12}^{(+)}$, and $E_{22}$. Figure 2(b) shows the diameter dependence of the energy difference $\Delta_{12}$ between $E_{12}^{(+)}$ and $E_{12}^{(-)}$ ($\Delta_{12} \equiv E_{12}^{(+)} - E_{12}^{(-)}$). $\Delta_{12}$ is approximately 200-300 meV for the observed nanotube species, and depends on the specific nanotube structure. The magnitude of the bright-dark energy splitting for transverse excitons is much larger than that of longitudinal excitons, as is consistent with previous theoretical predictions [21, 32].

Figure 3 shows PLE spectra of various (*n*, *m*) species for excitation perpendicular to the nanotube axis. Here we only plot (*n*, *m*) species for which we could observe both bright and dark transverse exciton absorption peaks in the observed excitation energy range. The higher and lower energy peaks are $E_{12}^{(+)}$ and $E_{12}^{(-)}$, respectively. We found strong (*n*, *m*) dependence of the $E_{12}^{(-)}$ peak intensities. Compared with the $E_{12}^{(+)}$ peak intensity $I_{12}^{(+)}$, near-zigzag SWNTs (close to (*n*, 0)) tend to have larger $E_{12}^{(-)}$ peak intensities $I_{12}^{(-)}$, while near-armchair SWNTs (close to (*n*, *n*)) tend to have smaller $I_{12}^{(-)}$.

In order to confirm our assignment of $E_{12}^{(-)}$ peaks to transverse *quasi* dark excitons and clarify the e-h asymmetry in SWNTs, we calculated absorption spectra of transverse excitons taking the depolarization effect and e-h asymmetry into account. Basically, we obtained the transverse exciton states in a $\mathbf{k} \cdot \mathbf{p}$ (or effective-mass) approximation by using a screened Hartree-Fock approximation [17]. So far, the e-h asymmetry has not been considered in the $\mathbf{k} \cdot \mathbf{p}$ approximation. Here, the overlap integral *s* causing the e-h asymmetry [7] is taken into account in the $\mathbf{k} \cdot \mathbf{p}$ approximation, and we include a higher-order term in the $\mathbf{k} \cdot \mathbf{p}$ Hamiltonian [33] in order to take the trigonal warping of energy bands into account. The trigonal warping provides the chirality dependence on exciton states. In addition to the overlap integral *s*, exciton states are determined by a hopping integral $\gamma_0$ and the dimensionless Coulomb interaction $v = (e^2/\kappa L)/(2\pi\gamma/L)$, where $\kappa$ is the effective dielectric constant of the SWNTs and $\gamma = (\sqrt{3}/2)a\gamma_0$ is a band parameter, where *a* is the lattice constant of graphene. In the following calculations we fixed $\gamma_0 = 2.6$ eV according to Ref. [30].

Figure 4(a) shows the calculated optical absorption spectra for (8, 6) SWNTs with ($s$=0.1) and without ($s$=0) e-h asymmetry. The nonradiative decay width was set to about 10 meV. For a finite value of the overlap integral, in addition to the bright exciton peak a small absorption peak emerges on the low-energy side due to the dark exciton that acquires oscillator strength because of the non-degenerate exciton wavefunctions in the K and K' valleys. This excellently reproduces the experimental observation in Fig. 3.

Figure 4(b) shows the bright-dark exciton energy splitting for $s$=0.1 as a function of Coulomb interaction parameter. We confirmed that the energy splitting is almost independent of $s$ for near-armchair SWNTs. From comparison with experimental data for (8, 6), (9, 5), and (10, 5) SWNTs, we determined the Coulomb interaction in SWNTs to be $v$=0.19, which is consistent with the estimation from the energy positions of $E_{11}$, $E_{22}$ and $E_{12}$ peaks [30].

Figure 4(c) shows the transverse dark exciton intensity normalized by the sum of bright and dark exciton intensities $I_{12}^{(-)}/(I_{12}^{(+)} + I_{12}^{(-)})$ as a function of the overlap integral $s$ for (8, 6), (9, 5) and (10, 5) SWNTs. In these cases, the trigonal warping and curvature effects are small and $s$ should be close to the value in graphene. Open symbols indicate experimental data. By comparing the calculated curves with the experimental data we can evaluate the overlap integral s~0.1 for these SWNTs. These values are close to the value $s$=0.129, which is conventionally used for graphene and SWNTs in simple tight-binding models [7].

In order to clarify the origin of the chirality dependence of $I_{12}^{(-)}$, we calculated the normalized intensities for various ($n$, $m$) SWNTs with $s$ from 0.08 to 0.22. Figure 4(d) shows $I_{12}^{(-)}/(I_{12}^{(+)} + I_{12}^{(-)})$ as a function of the energy difference between $E_{12}$ and $E_{21}$ exciton states ($E_{12} - E_{21}$). The $s$ value dominates the $E_{12} - E_{21}$ splitting, which becomes zero when the band structure is symmetric ($s$=0). Dashed lines in Fig. 4(d) indicate the experimental values for each ($n$, $m$). We found that the calculated results for various ($n$, $m$) species lie almost on the same curve when plotted as a function of $E_{12} - E_{21}$, while the $E_{12} - E_{21}$ values for the same $s$ tend to be larger for small diameter near-zigzag SWNTs. This suggests that the spectral weight transfer $I_{12}^{(-)}/(I_{12}^{(+)} + I_{12}^{(-)})$ is dominated by the degree of intrinsic degeneracy lifting between $E_{12}$ and $E_{21}$ exciton states. Comparison of experimental and theoretical results suggests that small diameter near-zigzag SWNTs tend to have large e-h asymmetry. This can be attributed to the strong trigonal warping effect that enhances the e-h asymmetry in SWNTs in small diameter near-zigzag species. The $s$ values that reproduce the experimental results of the near-zigzag SWNTs are ~0.2, which is considerably larger than that of near-armchair SWNTs ($s$~0.1). The larger $s$ values and stronger band asymmetry in near-zigzag SWNTs could be attributed to the enhanced band warping effect due to the contributions of second and third nearest-neighbor hoppings [8] that are not included in our calculations. Details of this effect on $I_{12}^{(-)}$ is beyond the scope of this study and remains as a future work.

In summary, we performed the first direct observation of the transverse quasi-dark exciton states brightened due to the intrinsic e-h asymmetry of SWNTs. In polarized PLE spectra we clearly observed a structure-dependent spectral weight transfer from transverse bright states to transverse dark states due to the degeneracy lifting of exciton wavefunctions. Based on comparison between our experimental and theoretical results, we evaluated the e-h asymmetry corresponding to the overlap integral $s$~0.1 in graphene and near-armchair SWNTs, and the strongly enhanced e-h asymmetry in small diameter near-zigzag SWNTs. Our findings complement the lack of fundamental information on the band asymmetry in graphene and SWNTs, and will lead to further understanding of these novel materials.

The authors are grateful to E. Einarsson (The University of Tokyo), T. Ando (Tokyo Institute of Technology), and S. Mazumdar (University of Arizona) for valuable discussions. Part of this work was financially supported by Grants-in-Aid for Scientific Research (19206024 and 19054003) from the Japan Society for the Promotion of Science, SCOPE (051403009) from the Ministry of Internal Affairs and Communications, and 'Development of Nanoelectronic Device Technology' of NEDO. One of the authors (YM) was financially supported by JSPS (No. 20-3712).

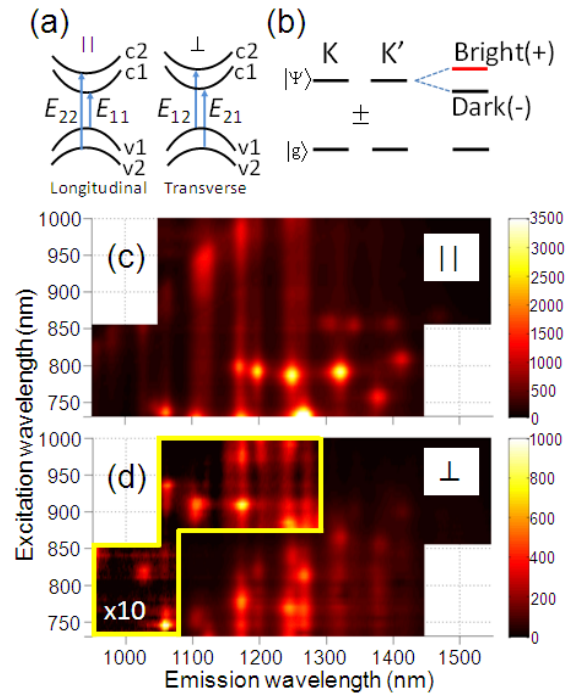

Figure 1. (color online)
Schematic diagram of (a) the selection rules for incident light polarized parallel (∥) and perpendicular (⊥) to the nanotube axis, and (b) intervalley mixing of K and K' excitons. The exciton wavefunctions are even and odd superpositions of those near the K and K' points in momentum space. This superposition gives the bright and dark states for longitudinal and transverse excitons. PLE maps for excitations polarized (c) parallel and (d) perpendicular to the nanotube axis. In (d), the PL intensities in the region surrounded by yellow lines have been magnified ten times.

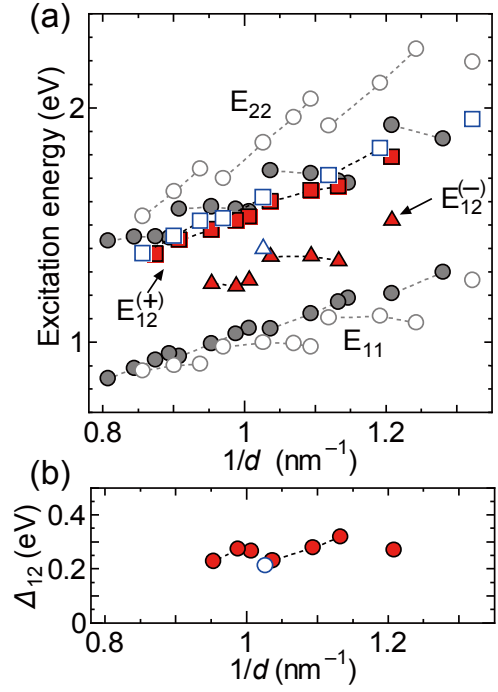

Figure 2. (color online)
(a) Excitation energy plot for $E_{11}$ and $E_{22}$ (circles), $E_{12}^{(+)}$ (squares), and $E_{12}^{(-)}$ (triangles) exciton states as a function of inverse diameter. $E_{12}^{(+)}$ for (7, 5) SWNTs were taken from Ref. [28]. $E_{11}$ and/or $E_{22}$ for small diameter SWNTs were obtained with Xe lamp excitation using a 5 nm slit width (excitation wavelength below 730 nm). (b) Energy difference between $E_{12}^{(+)}$ and $E_{12}^{(-)}$ plotted as a function of inverse diameter. In (a) and (b), filled and open marks correspond to type I ($2n+m$ mod 3 = 1) and type II SWNTs ($2n+m$ mod 3 = 2), respectively.

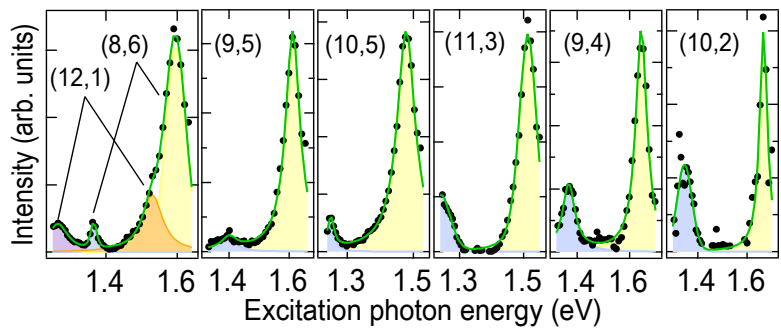

Figure. 3. (color online)
PLE spectra of various (*n*, *m*) species for excitation perpendicular to the nanotube axis. Higher and lower energy peaks correspond to $E_{12}^{(+)}$ and $E_{12}^{(-)}$, respectively. The spectra were decomposed by Voigt functions to evaluate each peak's area intensity. Since the PL emission wavelengths of (8, 6) and (12, 1) SWNTs are almost identical, the corresponding PLE spectra are combined.

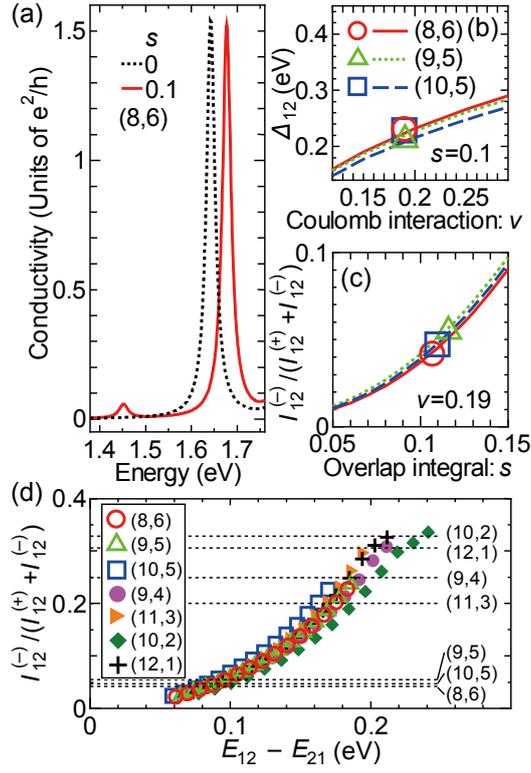

Figure 4. (color online)
(a) Calculated absorption spectra of an (8, 6) SWNT with ($s$=0.1) and without ($s$=0) e-h asymmetry. (b) The transverse exciton energy splitting $\Delta_{12}$ for (8, 6), (9, 5) and (10, 5) SWNTs as a function of Coulomb interaction parameter $v$. Experimental data are denoted by open symbols. (c) Calculated dark exciton intensities normalized by the sum of bright and dark exciton intensities as a function of the overlap integral $s$ for (8, 6), (9, 5) and (10, 5) SWNTs. Experimental data are denoted by open symbols. (d) Calculated normalized intensities of transverse dark excitons as a function of the energy difference between $E_{12}$ and $E_{21}$ transverse exciton states in the K and K' valleys. Experimental values are denoted by horizontal dashed lines.